\documentclass[journal]{IEEEtran}
\newcommand{\beq}{\begin{equation}}
\newcommand{\eeq}{\end{equation}}
\newcommand{\bsq}{\begin{subequations}}
\newcommand{\esq}{\end{subequations}}
\newcommand{\bq}{\begin{eqnarray}}
\newcommand{\eq}{\end{eqnarray}}
\newcommand{\bqn}{\begin{eqnarray*}}
\newcommand{\eqn}{\end{eqnarray*}}
\usepackage{cite}
\usepackage{makecell}
\usepackage{url}
\usepackage{empheq}
\usepackage{float}
\usepackage{dsfont}
\usepackage[hidelinks]{hyperref}

\usepackage{algorithm}
\usepackage{algorithmic}

\usepackage{amsmath,amssymb,amsfonts}
\allowdisplaybreaks[4]
\usepackage{graphicx}
\usepackage{textcomp}
\usepackage{amsmath}
\usepackage{enumerate}
\usepackage{epstopdf}
\usepackage{array}
\usepackage{booktabs}
\usepackage{subfigure}
\usepackage{multirow}
\usepackage{soul, color}
\usepackage[dvipsnames]{xcolor}
\usepackage[latin1]{inputenc}

\newtheorem{proposition}{Proposition}

\soulregister\cite7 
\soulregister\citep7 
\soulregister\citet7 
\soulregister\ref7
\soulregister\pageref7

\begin{document}
%\title{Review of Robust Optimization and Decision-Dependent Uncertainty}
\title{Robust Optimal Operation of Virtual Power Plants Under Decision-Dependent Uncertainty of \\ Price Elasticity}

\author{
Tao~Tan, 
Rui Xie,~\IEEEmembership{Member,~IEEE},
Meng Yang, and
Yue~Chen,~\IEEEmembership{Senior Member,~IEEE}

%\iffalse
\thanks{Corresponding author: Rui Xie}
% \thanks{T. Tan, R. Xie, M. Yang, and Y. Chen are with the Department of Mechanical and Automation Engineering, The Chinese University of Hong Kong, HKSAR, China (e-mail: ttan@mae.cuhk.edu.hk, ruixie@cuhk.edu.hk, myang@mae.cuhk.edu.hk, yuechen@mae.cuhk.edu.hk). }
\thanks{T. Tan, M. Yang, and Y. Chen are with the Department of Mechanical and Automation Engineering, The Chinese University of Hong Kong, HKSAR, China (e-mail: ttan@mae.cuhk.edu.hk, myang@mae.cuhk.edu.hk, yuechen@mae.cuhk.edu.hk). }
\thanks{R. Xie is with the Department of Mechanical and Automation Engineering, The Chinese University of Hong Kong, HKSAR, China, and also with the College of Electrical and Information Engineering, Hunan University, Changsha 410082, China (e-mail: xierui@hnu.edu.cn). }
}
\markboth{Journal of \LaTeX\ Class Files,~Vol.~XX, No.~X, Feb.~2025}%
{Shell \MakeLowercase{\textit{et al.}}: Bare Demo of IEEEtran.cls for IEEE Journals}
\maketitle

\begin{abstract}
The rapid deployment of distributed energy resources (DERs) is one of the essential efforts to mitigate global climate change. However, a vast number of small-scale DERs are difficult to manage individually, motivating the introduction of virtual power plants (VPPs). A VPP operator coordinates a group of DERs by setting suitable prices, and aggregates them for interaction with the power grid. In this context, optimal pricing plays a critical role in VPP operation. This paper proposes a robust optimal operation model for VPPs that considers uncertainty in the price elasticity of demand. Specifically, the demand elasticity is found to be influenced by the pricing decision, giving rise to decision-dependent uncertainty (DDU). An improved column-and-constraint (C\&CG) algorithm, together with tailored transformation and reformulation techniques, is developed to solve the robust model with DDU efficiently. Case studies based on actual electricity consumption data of London households demonstrate the effectiveness of the proposed model and algorithm. 
\end{abstract}


\section*{Nomenclature} %术语表，打星号是因为不用参与正文标题的标号排序

\subsection*{Acronym}
\begin{IEEEdescription}[\IEEEusemathlabelsep\IEEEsetlabelwidth{$i, j, x, y$}] % 设置符号的最大宽度
\item[C\&CG] Column-and-constraint generation.
\item[DDU] Decision-dependent uncertainty.
\item[DER] Distributed energy resource.
\item[DRO] Distributionally robust optimization.
\item[RO] Robust optimization.
\item[SP] Stochastic programming.
\item[TOU] Time-of-use.
\item[VPP] Virtual power plant.
\end{IEEEdescription}

\subsection*{Indices and Sets} % 小标题，可以分为集合，下标等等
%\addcontentsline{toc}{section}{Nomenclature}
\begin{IEEEdescription}[\IEEEusemathlabelsep\IEEEsetlabelwidth{$i, j, x, y$}] % 设置符号的最大宽度
\item[$\mathcal{I}$] Set of power nodes in VPP.
\item[$\mathcal{L}$] Set of power lines in VPP.
\item[$\mathcal{T}$] Set of periods.
\item[$\mathcal{K}$] Set of intervals of TOU price.
\end{IEEEdescription}

% 参数
\subsection*{Parameters}
\begin{IEEEdescription}[\IEEEusemathlabelsep\IEEEsetlabelwidth{$i, j, x, y$}] % 设置符号的最大宽度
\item[$L_{jt}$] Predicted load at node $j$ in period $t$.
\item[$\kappa_j$] Ratio of reactive demand to active demand at node $j$.
\item[$R_{i j}, X_{i j}$] Resistance and reactance of line $i \rightarrow j$.
\item[$\underline{P}_i, \overline{P}_i$] Lower and upper bounds of active power generation at node $i$.
\item[$\underline{Q}_i, \overline{Q}_i$] Lower and upper bounds of reactive power generation at node $i$.
\item[$\overline{S}_{i j}$] Maximum apparent power flow on line $i \rightarrow j$.
\item[$\underline{V}_i,\overline{V}_i$] Lower and upper bounds of the voltage magnitude at node $i$.
\item[$\underline{C}^{TOU}$] Lower bound of the TOU price.
\item[$\overline{C}^{TOU}$] Upper bound of the TOU price.
\item[$C_t^{REF}$] Reference price in period $t$.
\item[$r_{t k}^-,r_{t k}^+$] Lower and upper bounds of the ratio of TOU price to the reference price in period $t$ and interval $k$.
\item[$\rho_t$] Day-ahead market price in period $t$.
\item[$\rho_t^+,\rho_t^-$] Real-time purchasing/selling price from/to the power grid in period $t$.
\item[$\rho_i^G$] Unit generation cost of active power at node $i$.
% \item[$\rho^{SHED}$] Unit load shedding cost.
% \item[$\rho^{CUR}$] Unit load curtailment cost.

\item[$\xi_{i t k}^-, \xi_{i t k}^+$] Lower and upper bounds of price elasticity at node $i$ in period $t$ and interval $k$.
\end{IEEEdescription}

\subsection*{Variables}

\subsubsection*{First-Stage (Day-Ahead) Variables}

\begin{IEEEdescription}[\IEEEusemathlabelsep\IEEEsetlabelwidth{$i, j, x, y$}] % 设置符号的最大宽度
\item[$p_{i t}^{DA}, q_{i t}^{DA}$] Scheduled active and reactive power generation at node $i$ in period $t$.
\item[$p_{i j t}^{DA}, q_{i j t}^{DA}$] Scheduled inflow active and reactive power on line $i \rightarrow j$ in period $t$.
\item[$v_{i t}^{DA}$] Square of scheduled voltage magnitude at node $i$ in period $t$.
\item[$c_t^{TOU}$] TOU price in period $t$.
\item[$p_{0 t}^{DA}$] Scheduled active power injection at node $0$ in period $t$.
\end{IEEEdescription}	

\subsubsection*{Uncertainty Variables}

\begin{IEEEdescription}[\IEEEusemathlabelsep\IEEEsetlabelwidth{$i, j, x, y$}] % 设置符号的最大宽度
\item[$\xi_{i t}$] Price elasticity at node $i$ in period $t$.
\end{IEEEdescription}

\subsubsection*{Second-Stage (Real-Time) Variables}

\begin{IEEEdescription}[\IEEEusemathlabelsep\IEEEsetlabelwidth{$i, j, x, y$}] % 设置符号的最大宽度
\item[$l_{i t}$] Real-time demand at node $i$ in period $t$.
\item[$p_{i t}^{RT}, q_{i t}^{RT}$] Real-time active and reactive power at node $i$ in period $t$.
\item[$p_{i j t}^{RT}, q_{i j t}^{RT}$] Real-time inflow active and reactive power on line $i \rightarrow j$ in period $t$.
\item[$v_{i t}^{RT}$] Square of real-time voltage magnitude at node $i$ in period $t$.
\item[$p_{0 t}^{\Delta +}, p_{0 t}^{\Delta -}$] Upward and downward power adjustment at node $0$ in period $t$.
% \item[$r_{i t}^+$] Load shedding power at node $i$ in period $t$. 
% \item[$r_{i t}^-$] Load curtailment power at node $i$ in period $t$. 
\end{IEEEdescription}	

\section{Introduction}
%VPP uncertainty
%RO
%Decision

Distributed energy resources (DERs), such as electric vehicles, battery energy storage, and rooftop solar panels, have been widely deployed to reduce dependence of power grids on fossil fuels and mitigate global warming \cite{IEA2025}. In this context, virtual power plants (VPP) have been introduced, which aggregate multiple small-scale DERs through intelligent control to function collectively like traditional large power plants \cite{ruan2024VPP}. During the operation of a VPP, the internal pricing strategies to incentivize the participation of DERs are important, which need to consider the uncertainty of DER response \cite{ullah2019comprehensive}.

%As renewable energy continues to increase \cite{IEA2025}, the concept of virtual power plant (VPP) consisting of renewable energy and intelligent control \cite{ruan2024VPP} is drawing increasing attention. During the operation of the VPP, the fluctuation and intermittency of renewable energy \cite{ullah2019comprehensive} and the variance in consumer demand add uncertainties to the system, which is a threat to the security and reliability of power systems.   

To deal with uncertainty in VPPs, three typical methodologies are widely adopted, including stochastic programming (SP), robust optimization (RO), and distributionally robust optimization (DRO). SP models uncertain factors, e.g., renewable power generation, as random variables with given probability distributions, and derives the optimal solution based on sampling or chance constraints \cite{shinde2022multistage}. Although SP is straightforward, it requires exact and detailed probability distributions, which makes it less practical. Moreover, it is computationally demanding to solve large-scale SP problems with numerous scenarios. RO aims to determine a solution that remains feasible and optimal under a range of possible scenarios \cite{Wulro}. It is easier to implement than SP without requiring detailed probabilistic information. DRO extends RO by optimizing the decision over a set of probability distributions rather than a set of scenarios \cite{yu2023flexible}. Although DRO can obtain a less conservative result than RO, its formulation and solution are significantly more complex. More importantly, SP and DRO allow a certain level of risk \cite{cao2023distributionally}, making it difficult to guarantee 100\% confidence in the security of the VPP. For these reasons, RO is a more suitable method to optimize the operation of VPPs where security is of top priority. 

RO has been widely employed in VPPs. For example, the day-ahead energy storage scheduling within a VPP was studied in \cite{xiang2023optimizing}, considering the uncertainty of renewable generation and load. In \cite{nemati2025assessingvaluerenewablebasedvpp}, the values of renewable-only VPPs and grid-scale energy storage systems in energy and reserve markets were compared using a two-stage RO framework. These studies mainly focus on the decision-independent uncertainty (DIU), in which the decisions in the first stage have no impact on the uncertainty set. 
RO with DIU can be solved efficiently using the Benders decomposition \cite{bertsimas2012adaptive}, the column-and-constraint generation (C\&CG) algorithm \cite{zeng2013solving}, etc. These algorithms iteratively identify the worst-case scenarios and return the scenarios or the corresponding cutting planes to the first-stage problem until convergence. 

Recently, decision-dependent uncertainty (DDU), where the first-stage decisions can affect the uncertainty set, has been widely recognized. RO with DDU is an emerging topic in power systems \cite{tan2025adjustable} and there are few studies considering DDU in VPPs. A stochastic adaptive RO model for a VPP's scheduling in day-ahead energy-reserve markets was established in \cite{zhang2021robust}, considering the DDU of real-time reserve deployment requests. The DDU from energy storage operation was modeled in \cite{qi2023portfolio} and then integrated into a stochastic optimization method for VPP.
% RO with DDU is an emerging topic in power systems \cite{chen2016robust,zhangRES,tan2025adjustable} and there are few studies considering DDU in the VPP system \cite{zhang2021robust,qi2023portfolio}.

DDU substantially adds difficulties to modeling RO problems, because decisions can change the uncertainty set in various ways \cite{hellemo2018decision}. 
% The uncertain load demand impact by first-stage time-of-use (TOU) electricity prices studied in this paper is one of the DDU examples. In this case, previous modeling and solution methods may not be applicable in our study. 
%As the interdependence between uncertainty and first-stage decisions becomes more prevalent, an increasing number of decision-dependent uncertainties (DDU) in power systems have been noted \cite{chen2016robust,zhangRES,tan2025adjustable} 
%Also, there are a few studies considering DDU in the VPP system \cite{zhang2021robust,qi2023portfolio}. 
%DDU shuoyixia buyiyang//alg why our integer uncertainty set, nonlinear//conclude before contribution lengthen
%Decisions in the first stage can affect the probability distributions of uncertainties, the dimension of uncertainty factors, or when uncertainty realizes \cite{hellemo2018decision}. 
A typical example is demand response, a popular research topic in VPPs, in which studies encompass optimal scheduling \cite{zhou2023urban}, bidding strategy \cite{mei2023optimal}, and real-time pricing \cite{kong2023real}.  
The responsiveness of electricity demand is influenced by time-of-use (TOU) pricing \cite{lu2025assessing}. Such responsiveness is usually characterized by a concept called price elasticity \cite{kirschen2004fundamentals}. In this paper, the uncertainty of price elasticity and how it is influenced by the pricing decision are considered.
%{\color{red}When considering DDU, DR in the VPP system is easily influenced by decision variables such as electricity price set by VPP operators, the availability of flexible loads, the scale of storage system, etc. Among these factors, the TOU electricity price is a rather convenient way to guide consumers to shift load \cite{lu2025assessing}. The change in the price of electricity will have an impact on demand \cite{kirschen2004fundamentals, conejo2010real,thimmapuram2013consumers}. For these reasons, we planned to develop a model that sets the electricity rate as a decision variable and includes DR in the VPP system.} 

In terms of solving RO problems with DDU, traditional methods are found to be intractable, suboptimal, or lacking convergence guarantees \cite{chen2022robust}. Therefore, innovative modeling techniques and solution algorithms are required to address RO with DDU. Currently, several variants of C\&CG have been developed for this problem. Adaptive C\&CG is an example, which returns the set of active constraints to represent the worst-case scenario identified \cite{chen2022robust}. Another algorithm is mapping C\&CG, which applies mapping rules to maintain the worst-case scenarios at the vertices of the new uncertainty set \cite{Xieload}. Dual C\&CG and transformation-based C\&CG apply duality and variable substitution to convert an RO with DDU problem to an RO with DIU problem \cite{tan2024robust, yang2025robust}. There are also some other algorithms, including variants of Benders decomposition \cite{zhang2022two}, multi-parametric programming \cite{avraamidou2020adjustable}, K-adaptability \cite{vayanos2025robust}, and generic solution algorithms \cite{zhang2025generic}. However, these general algorithms might be time-consuming and difficult to implement due to the complicated transformation and projection.

% {\color{red}We planned to develop a new algorithm to solve the RO with DDU problem in VPP. In our model, previous algorithms may not be solvable because there is a bilinear part in the objective function and integers in the uncertainty set.}
%lengthen

This paper aims to fill the research gaps above by proposing a robust VPP optimal operation model considering the impact of pricing strategies on the uncertain demand elasticity, and developing a novel and tractable solution algorithm based on the special structure of the problem studied. Our main contribution is two-fold:

\begin{enumerate}
    \item \emph{Robust VPP Optimal Operation Model Under DDU}. We propose a two-stage robust optimal operation model for VPPs to determine their internal TOU tariffs for minimizing the overall cost. Particularly, the impact of first-stage pricing decisions on the uncertain price elasticity is considered, which renders DDU. Compared to traditional robust VPP operation models with DIU only, the proposed model accounts for the mutual impact between decisions and uncertainty, and thus gives more precise optimal operation and pricing decisions.
    \item \emph{Improved C\&CG Algorithm}. To solve the robust optimal operation problem of a VPP with DDU, we develop an improved C\&CG algorithm that leverages an uncertainty-set transformation. Compared with standard RO solution methods that may suffer from non-convergence or suboptimality, the proposed algorithm is guaranteed to converge to a globally optimal solution.
\end{enumerate}

%1) We developed a two-stage robust dispatch model to determine TOU tariff and the generation resources. The uncertain demand is decision-dependent on the TOU price, which is derived from a clustering algorithm. 

%2): We customized an improved C\&CG algorithm. This algorithm can solve the problem in the model while traditional C\&CG cannot.
%In this model, the TOU is determined on the first stage, and the demand will change so there will be a re-dispatch process on the second stage. 

The remainder of the paper is organized as follows. 
Section~\ref{sec:model} introduces the robust optimal operation model for VPPs considering DDU.
The solution methodology is presented in Section~\ref{sec:methodology}.
Numerical case studies are conducted in Section~\ref{sec:case}. 
Finally, conclusions are summarized in Section~\ref{sec:conclusion}.

%\section{Modeling}

\section{Robust Optimal Operation Model of Virtual Power Plants}
\label{sec:model}

\begin{figure}[ht]
\centering
\includegraphics[width=1.0\columnwidth]{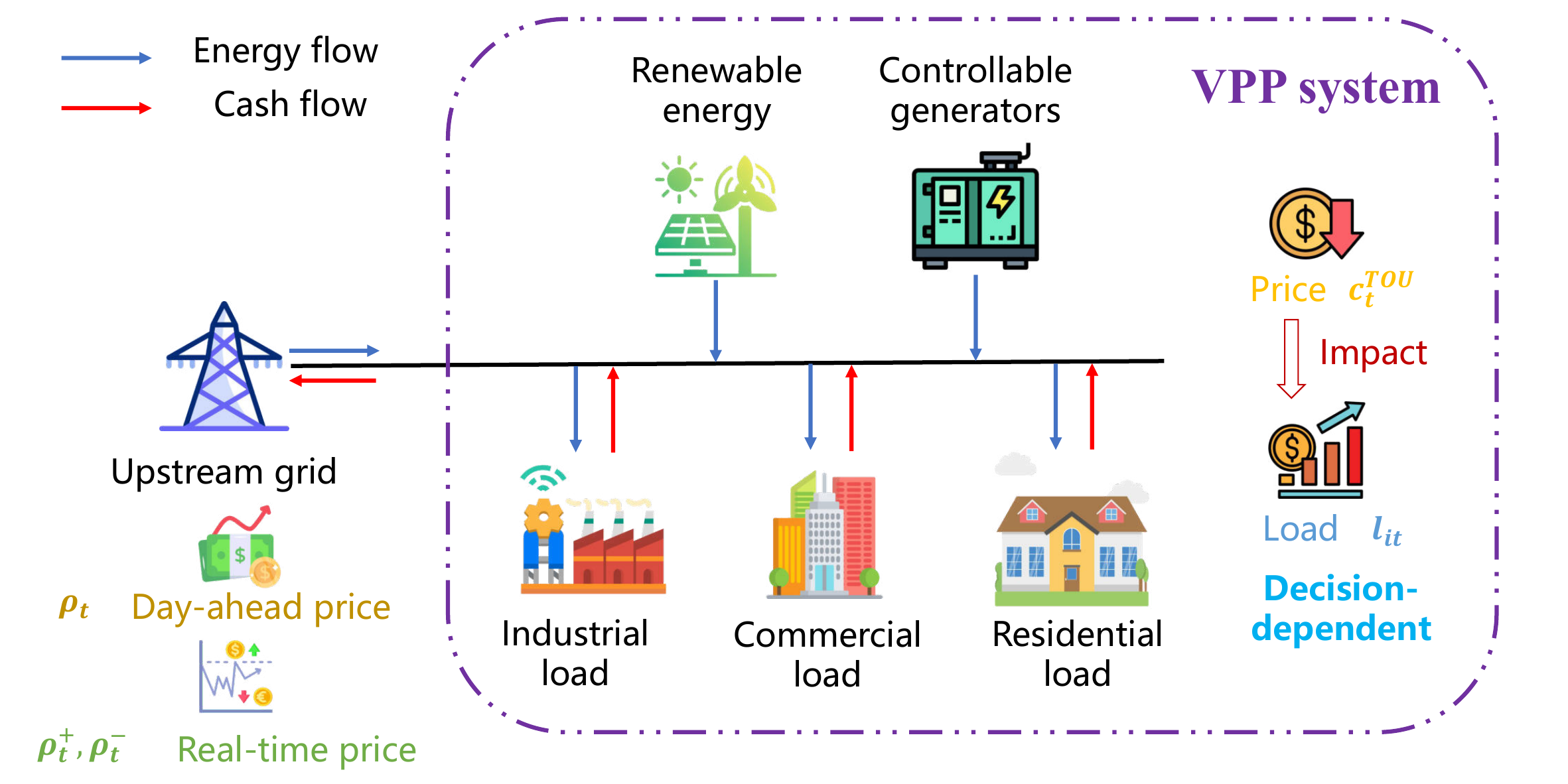}
\caption{The overview of the VPP system.}
\label{fig:overview}
\end{figure}

In this section, we propose a robust optimal operation model for a VPP, considering the impact of pricing decisions on the demand uncertainty. The overview of the VPP system is shown in Fig. \ref{fig:overview}. We first introduce the first-stage constraints, the decision-dependent uncertainty set, and the second-stage constraints, respectively. Then, we give the overall robust optimal operation model.

\subsection{First-Stage Constraints}
In the first stage, the VPP operator determines the day-ahead setpoints for generators and the TOU prices for demand, subject to power network constraints. Since the VPP resides at the distribution level, the LinDistFlow model is adopted. Particularly, the first-stage operational constraints include the following:
\begin{subequations}
\label{eq:first-stage}
\begin{align}
& p_{i j t}^{DA} + p_{j t}^{DA} - L_{j t} = \sum_{l: j \rightarrow l} p_{j l t}^{DA}, \forall i \rightarrow j \in \mathcal{L}, \forall t \in \mathcal{T}, \label{eq:first-stage-1}\\
& q_{i j t}^{DA} + q_{j t}^{DA} - \kappa_j L_{j t} = \sum_{l: j \rightarrow l} q_{j l t}^{DA}, \forall i \rightarrow j \in \mathcal{L}, \forall t \in \mathcal{T}, \label{eq:first-stage-2}\\
& v_{j t}^{DA} = v_{i t}^{DA} - 2 R_{i j} p_{i j t}^{DA} - 2 X_{i j} q_{i j t}^{DA}, \forall i \rightarrow j \in \mathcal{L}, \forall t \in \mathcal{T}, \label{eq:first-stage-3}\\
& \underline{P}_i \leq p_{i t}^{DA} \leq \overline{P}_i, \underline{Q}_i \leq q_{i t}^{DA} \leq \overline{Q}_i, \forall i \in \mathcal{I}, \forall t \in \mathcal{T}, \label{eq:first-stage-4}\\
& (p_{i j t}^{DA})^2 + (q_{i j t}^{DA})^2 \leq (\overline{S}_{i j})^2, \forall i \rightarrow j \in \mathcal{L}, \forall t \in \mathcal{T}, \label{eq:first-stage-5}\\
& (\underline{V}_i)^2 \leq v_{i t}^{DA} \leq (\overline{V}_i)^2, \forall i \in \mathcal{I}, \forall t \in \mathcal{T}, \label{eq:first-stage-6}\\
& \underline{C}^{TOU} \leq c_t^{TOU} \leq \overline{C}^{TOU}, \forall t \in \mathcal{T}. \label{eq:first-stage-7}
\end{align}
\end{subequations}
Constraints \eqref{eq:first-stage-1} and \eqref{eq:first-stage-2} describe the active and reactive power balance at node $j \in \mathcal{I}$ in period $t \in \mathcal{T}$. Here, $p_{j t}^{DA}$ and $q_{j t}^{DA}$ are the scheduled active and reactive power generation, $p_{i j t}^{DA}$ and $q_{i j t}^{DA}$ are the scheduled inflow active and reactive power on line $i \rightarrow j$, $L_{j t}$ is the predicted active load, and $\kappa_j$ is the ratio of reactive to active demand. Constraint \eqref{eq:first-stage-3} models the voltage drop along power line $i \rightarrow j$, where $v_{i t}^{DA}$ is the square of the voltage magnitude at node $i$, and $R_{ij}$ and $X_{ij}$ are the line resistance and reactance, respectively. Constraint \eqref{eq:first-stage-4} enforces the generation limits for active and reactive power at node $i$. Constraint \eqref{eq:first-stage-5} limits the apparent power flow on line $i \rightarrow j$ to its maximum rating $\overline{S}_{i j}$. Constraint \eqref{eq:first-stage-6} ensures that the square of the voltage magnitude at each node remains within its secure range. Finally, constraint \eqref{eq:first-stage-7} bounds the TOU price $c_t^{TOU}$.

\subsection{Decision-Dependent Uncertainty Set}
In this paper, we focus on the demand uncertainty; or more specifically, the uncertain price elasticity $\xi_{it}$ indicating how demand changes with prices. Based on analysis of real-world data (given in Section \ref{sec:case} later), we find that the variation range of $\xi_{it}$ depends on the ratio between the determined TOU price $c_t^{TOU}$ and the reference price $C_t^{REF}$. Therefore, the uncertainty set is formulated as a function of the first-stage decision vector $c^{TOU} = (c_t^{TOU})_{t\in\mathcal{T}}$. We denote this set as $\mathcal{U}(c^{TOU})$:
\begin{align}
\mathcal{U}(c^{TOU}) = \left\{ \xi = (\xi_{i t})_{i\in\mathcal{I}, t\in\mathcal{T}} \middle|
\begin{array}{l}
\forall i \in \mathcal{I}, \forall t \in \mathcal{T}, \forall k \in \mathcal{K}: \\
\text{if } r_{t k}^- \leq \frac{c_t^{TOU}}{C_t^{REF}} \leq r_{t k}^+, \\
\text{then } \xi_{i t k}^- \leq \xi_{i t} \leq \xi_{i t k}^+
\end{array}
\right\}.
\label{eq:uncertainty-set-origin}
\end{align}
In \eqref{eq:uncertainty-set-origin}, the set of intervals $\mathcal{K}$ partitions the possible price ratios. For interval $k \in \mathcal{K}$ defined by bounds $[r_{t k}^-, r_{t k}^+]$, there is a corresponding set of elasticity bounds $[\xi_{i t k}^-, \xi_{i t k}^+]$ for the uncertainty variable $\xi_{i t}$. 
This uncertainty set is essentially a decision-dependent uncertainty set because the first-stage decision $c_t^{TOU}$ influences the range of $\xi_{it}$. This decision dependency challenges the solution of the RO problem, as revealed by \cite{chen2022robust}. In Section \ref{sec:methodology}, we propose a reformulation of the uncertainty set to address this challenge.

\subsection{Second-Stage Constraints}
In the second stage, after observing the actual demand, the VPP operator adjusts the power output of generators to maintain real-time balance between electricity supply and demand. The second-stage constraints include the following:
\begin{subequations}
\label{eq:second-stage}
\begin{align}
& l_{i t} = L_{i t}(1 + \xi_{i t}(c_t^{TOU} / C_t^{REF} - 1)), \forall i \in \mathcal{I},
\forall t \in \mathcal{T}, \label{eq:second-stage-1}\\
& p_{0 t}^{DA} + p_{0 t}^{\Delta +} - p_{0 t}^{\Delta -} = p_{0 t}^{RT}, \forall t \in \mathcal{T}, \label{eq:second-stage-2a}\\
& p_{0 t}^{\Delta +} \geq 0, p_{0 t}^{\Delta -} \geq 0, \forall t \in \mathcal{T}, \label{eq:second-stage-2b}\\
& p_{i j t}^{RT} + p_{j t}^{RT} - l_{j t} = \sum_{l: j \rightarrow l} p_{j l t}^{RT}, \forall i \rightarrow j \in \mathcal{L}, \forall t \in \mathcal{T}, \label{eq:second-stage-4}\\
& q_{i j t}^{RT} + q_{j t}^{RT} - \kappa_j l_{j t} = \sum_{l: j \rightarrow l} q_{j l t}^{RT}, \forall i \rightarrow j \in \mathcal{L}, \forall t \in \mathcal{T}, \label{eq:second-stage-5}\\
& v_{j t}^{RT} = v_{i t}^{RT} - 2 R_{i j} p_{i j t}^{RT} - 2 X_{i j} q_{i j t}^{RT}, \forall i \rightarrow j \in \mathcal{L}, \forall t \in \mathcal{T}, \label{eq:second-stage-6}\\
& \underline{P}_i \leq p_{i t}^{RT} \leq \overline{P}_i, \underline{Q}_i \leq q_{i t}^{RT} \leq \overline{Q}_i, \forall i \in \mathcal{I}, \forall t \in \mathcal{T}, \label{eq:second-stage-7}\\
& (p_{i j t}^{RT})^2 + (q_{i j t}^{RT})^2 \leq (\overline{S}_{i j})^2, \forall i \rightarrow j \in \mathcal{L}, \forall t \in \mathcal{T}, \label{eq:second-stage-8}\\
& (\underline{V}_i)^2 \leq v_{i t}^{RT} \leq (\overline{V}_i)^2, \forall i \in \mathcal{I}, \forall t \in \mathcal{T}. \label{eq:second-stage-9}
\end{align}
\end{subequations}
Constraint \eqref{eq:second-stage-1} defines the realized demand $l_{it}$ based on the predicted load $L_{it}$, the TOU price $c_t^{TOU}$, and the realized uncertainty $\xi_{it}$. Constraints \eqref{eq:second-stage-2a} and \eqref{eq:second-stage-2b} model the real-time power $p_{0 t}^{RT}$ purchased from the upstream grid. This is composed of the day-ahead schedule $p_{0 t}^{DA}$ and upward $p_{0 t}^{\Delta +}$ or downward $p_{0 t}^{\Delta -}$ adjustments. 
% Constraint \eqref{eq:second-stage-3} ensures that load shedding and curtailment $r_{i t}^+,r_{i t}^-$ is nonnegative. 
Constraints \eqref{eq:second-stage-4} and \eqref{eq:second-stage-5} enforce the real-time active and reactive power balance, respectively. The remaining constraints \eqref{eq:second-stage-6}--\eqref{eq:second-stage-9} are the real-time network operational constraints. They model the voltage drop, enforce generation limits, bound the line apparent power, and ensure that the square of the voltage magnitude remains within its secure range.

\subsection{Overall Robust Optimal Operation Model}
The overall robust optimal operation model for the VPP is formulated as a two-stage RO problem. We use notation $x$ to denote the collection of the first-stage decision variables $p_{i t}^{DA}$, $q_{i t}^{DA}$, $p_{i j t}^{DA}$, $q_{i j t}^{DA}$, $v_{i t}^{DA}$, and $c_t^{TOU}$, and $y$ for the collection of second-stage variables $l_{i t}$, $p_{i t}^{RT}$, $q_{i t}^{RT}$, $p_{i j t}^{RT}$, $q_{i j t}^{RT}$, $v_{i t}^{RT}$, $p_{0 t}^{\Delta +}$, and $p_{0 t}^{\Delta -}$. The model is as follows:
% \begin{align}
% \label{eq:RO}
% & \min_{x \in \mathcal{X}} \Bigg\{ \sum_{t \in \mathcal{T}} \rho_t p_{0 t}^{DA} + \max_{\xi \in \mathcal{U}(c^{TOU})} \min_{y \in \mathcal{Y}(x, \xi)} \nonumber\\
% & \Bigg\{ - \sum_{t \in \mathcal{T}} \sum_{i \in \mathcal{I}} c_t^{TOU} l_{i t} + \sum_{t \in \mathcal{T}} \left( \rho_t^+ p_{0 t}^{\Delta +} - \rho_t^- p_{0 t}^{\Delta -} \right) \nonumber\\
% & + \sum_{t \in \mathcal{T}} \sum_{i \in \mathcal{I}} \rho_i^G p_{i t}^{RT} + \sum_{t \in \mathcal{T}} \sum_{i \in \mathcal{I}} (\rho^{SHED} r_{i t}^++\rho^{CUR} r_{i t}^-) \Bigg\} \Bigg\},
% \end{align}
\begin{align}
\label{eq:RO}
& \min_{x \in \mathcal{X}} \Bigg\{ \sum_{t \in \mathcal{T}} \rho_t p_{0 t}^{DA} + \max_{\xi \in \mathcal{U}(c^{TOU})} \min_{y \in \mathcal{Y}(x, \xi)} \nonumber\\
& \Bigg\{ - \sum_{t \in \mathcal{T}} \sum_{i \in \mathcal{I}} c_t^{TOU} l_{i t} + \sum_{t \in \mathcal{T}} \left( \rho_t^+ p_{0 t}^{\Delta +} - \rho_t^- p_{0 t}^{\Delta -} \right) \nonumber\\
& + \sum_{t \in \mathcal{T}} \sum_{i \in \mathcal{I}} \rho_i^G p_{i t}^{RT} \Bigg\} \Bigg\},
\end{align}
where the uncertainty set $\mathcal{U}(c^{TOU})$ is defined in \eqref{eq:uncertainty-set-origin}, and the first- and second-stage feasible regions ($\mathcal{X}$ and $\mathcal{Y}(x, \xi)$) are:
% \begin{subequations}
% \begin{align}
%     & \mathcal{X} = \left\{ x = (p_{i t}^{DA}, q_{i t}^{DA}, p_{i j t}^{DA}, q_{i j t}^{DA}, v_{i t}^{DA}, c_t^{TOU}) ~\middle|~ \eqref{eq:first-stage} \right\}, \\
%     & \mathcal{Y}(x, \xi) = \nonumber \\
%     & \left\{ y = (l_{i t}, p_{i t}^{RT}, q_{i t}^{RT}, p_{i j t}^{RT}, q_{i j t}^{RT}, v_{i t}^{RT}, p_{0 t}^{\Delta +}, p_{0 t}^{\Delta -}, r_{i t}^+,r_{i t}^-) ~\middle|~ \eqref{eq:second-stage} \right\}.
% \end{align}
% \end{subequations}
\begin{subequations}
\begin{align}
    & \mathcal{X} = \left\{ x = (p_{i t}^{DA}, q_{i t}^{DA}, p_{i j t}^{DA}, q_{i j t}^{DA}, v_{i t}^{DA}, c_t^{TOU}) ~\middle|~ \eqref{eq:first-stage} \right\}, \\
    & \mathcal{Y}(x, \xi) = \nonumber \\
    & \left\{ y = (l_{i t}, p_{i t}^{RT}, q_{i t}^{RT}, p_{i j t}^{RT}, q_{i j t}^{RT}, v_{i t}^{RT}, p_{0 t}^{\Delta +}, p_{0 t}^{\Delta -}) ~\middle|~ \eqref{eq:second-stage} \right\}.
\end{align}
\end{subequations}

The objective function in \eqref{eq:RO} is composed of two parts. The first term $\sum_{t \in \mathcal{T}} \rho_t p_{0 t}^{DA}$ is the first-stage (day-ahead) cost of purchasing energy from the upstream grid at day-ahead price $\rho_t$. The second part is the worst-case second-stage (real-time) operational cost. The max operator finds the worst-case scenario of the price elasticity $\xi$ from the decision-dependent uncertainty set $\mathcal{U}(c^{TOU})$. The inner min operator represents the VPP's real-time adjustments to minimize total operational costs, including the negative income of selling electricity to clients $\sum_{t \in \mathcal{T}} \sum_{i \in \mathcal{I}} c_t^{TOU} l_{i t}$, the adjustment cost $\sum_{t \in \mathcal{T}} ( \rho_t^+ p_{0 t}^{\Delta +} - \rho_t^- p_{0 t}^{\Delta -})$ for purchasing/selling power from/to the upstream grid based on the real-time prices, and the actual power generation cost $\sum_{t \in \mathcal{T}} \sum_{i \in \mathcal{I}} \rho_i^G p_{i t}^{RT}$.
% and the load shedding and curtailment cost $\sum_{t \in \mathcal{T}} \sum_{i \in \mathcal{I}} (\rho^{SHED} r_{i t}^++\rho^{CUR} r_{i t}^-) $.

\section{Solution Method}
\label{sec:methodology}

Solving the robust VPP optimal operation model \eqref{eq:RO} faces two challenges: (i) The nonlinear constraint \eqref{eq:second-stage-8} hinders the transformation of the second-stage problem based on optimality conditions. (ii) The decision-dependent uncertainty set \eqref{eq:uncertainty-set-origin} makes traditional RO algorithms inapplicable.

\subsection{Linearization of Quadratic Constraints}
\label{sec:III-A}
We first address the challenge (i). The constraint $(p_{i j t}^{RT})^2 + (q_{i j t}^{RT})^2 \leq (\overline{S}_{i j})^2$ represents a disk region on the $(p_{i j t}^{RT}, q_{i j t}^{RT})$ plane, which can be approximated by its inscribed regular polygons with $S$ sides. Therefore, we use the following constraints to inner approximate $(p_{i j t}^{RT})^2 + (q_{i j t}^{RT})^2 \leq (\overline{S}_{i j})^2$:
\begin{align}
    & p_{i j t}^{RT} \cos \left( \frac{2 s - 1}{S} \pi \right) + q_{i j t}^{RT} \sin \left( \frac{2 s - 1}{S} \pi \right)  \leq \cos \left( \frac{\pi}{S} \right) \overline{S}_{i j}, \nonumber\\
    & \forall s = 1, 2, \dots, S.
\end{align}

\begin{figure}[ht]
\centering
\includegraphics[width=0.9\columnwidth]{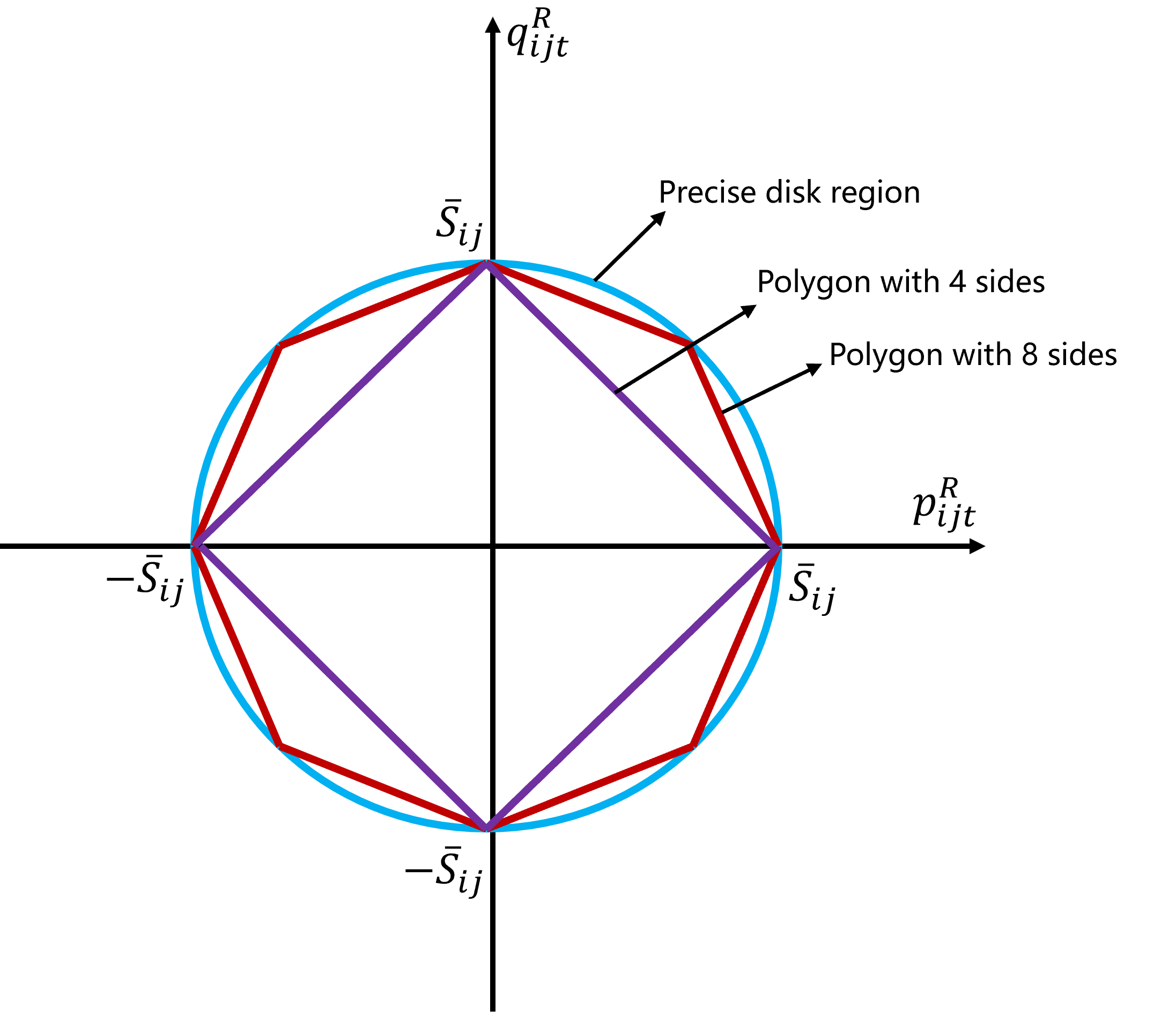}
\caption{The basic idea of inner approximation.}
\label{fig:approximation}
\end{figure}

Fig. \ref{fig:approximation} illustrates the basic idea of this inner approximation. As we can see, the larger the $S$, the more accurate the approximation. After this transformation, the second-stage problem turns out to be a linear program (LP), which facilitates the subsequent transformation.

\subsection{Uncertainty Set Reformulation}
Next, to address challenge (ii), we propose an equivalent transformation \eqref{eq:uncertainty-set} of the uncertainty set \eqref{eq:uncertainty-set-origin}. The binary indicators $z_{tk}$ identify the interval $[r_{tk}^{-}, r_{tk}^{+}]$ into which the TOU-to-reference price ratio $c_t^{TOU}/c^{REF}$ falls and accordingly determine the corresponding price elasticity $\xi_{it}$.
\begin{subequations}
\label{eq:uncertainty-set}
\begin{align}
    \label{eq:uncertainty-set-1}
    & \sum_{k \in \mathcal{K}} \xi_{i t k}^- z_{t k} \leq \xi_{i t} \leq \sum_{k \in \mathcal{K}} \xi_{i t k}^+ z_{t k}, \forall i \in \mathcal{I}, \forall t \in \mathcal{T}, \\
    \label{eq:uncertainty-set-2}
    & \sum_{k \in \mathcal{K}} r_{t k}^- z_{t k} \leq \frac{c_t^{TOU}}{c^{REF}} \leq \sum_{k \in \mathcal{K}} r_{t k}^+ z_{t k}, \forall t \in \mathcal{T}, \\
    \label{eq:uncertainty-set-3}
    & \sum_{k \in \mathcal{K}} z_{t k} = 1, \forall t \in \mathcal{T}, \\
    \label{eq:uncertainty-set-4}
    & z_{t k} \in \{0, 1\}, \forall k \in \mathcal{K}, \forall t \in \mathcal{T}.
\end{align}
\end{subequations}

Specifically, constraints \eqref{eq:uncertainty-set-3} and \eqref{eq:uncertainty-set-4} ensure that for each time period $t$ there is only one $z_{tk}=1$ while all others equal 0. Suppose $z_{t\bar k}=1$, then constraint \eqref{eq:uncertainty-set-2}  gives $r_{t\bar k}^{-} \le \frac{c_t^{TOU}}{c^{REF}} \le  r_{t\bar k}^+$, and correspondingly \eqref{eq:uncertainty-set-1} gives $\xi_{it\bar k}^{-} \le \xi_{it} \le \xi_{it\bar k}^{+}$, which is consistent with \eqref{eq:uncertainty-set-origin}.

%{\color{red} add sth about the uncertainty budget}

\subsection{Improved C\&CG Algorithm}
After the above transformation and reformulation, the robust optimal operation model \eqref{eq:RO} can be restructured in a compact form as shown in \eqref{compact}:
\begin{subequations}\label{compact}
\begin{align}
\min_{x, z}~ & C^\top x + \max_\xi \min_y E(x)^\top y ,\\
\text{s.t.}~ & A x \geq B, \eqref{eq:uncertainty-set}, P y \geq Q(x) \xi + R(x),
\end{align}
\end{subequations}
where $x$ and $y$ collect all the first- and second-stage decision variables, respectively; $A$, $B$, $C$, and $P$ are the corresponding coefficient matrices; the coefficient matrices $E(x)$, $Q(x)$, and $R(x)$ are influenced by the first-stage decision $x$. In the following, we introduce our improved C\&CG algorithm to solve the problem, which iteratively solves a master problem to update the first-stage decision $x$ and solves a feasibility check problem or a subproblem to identify the worst-case scenario under given $x$.

\subsubsection{Feasibility Check Problem and Subproblem}
First, given the first-stage decision $x$, we formulate the feasibility check problem (or subproblem) to identify the worst-case scenario that causes infeasibility (or the highest cost). The feasibility check problem is formulated as follows:
\begin{subequations}
\label{eq:FC}
\begin{align}
\textbf{FC}: ~\max_\xi~ & \min_{y,s} 1^\top s, \\
\text{s.t.}~ & \eqref{eq:uncertainty-set}, P y + s \geq Q(x) \xi + R(x), s \geq 0.
\end{align}
\end{subequations}
The \textbf{FC} problem is always feasible since $y=0$ and $s=\max\{0, Q(x)\xi+R(x)\}$ is always a feasible solution. Moreover, if the optimal solution to \eqref{eq:FC} satisfies $s^*=0$, then for every $\xi$ in the decision-dependent uncertainty set \eqref{eq:uncertainty-set}, there exists a feasible solution satisfying the original constraint $Py \ge Q(x)\xi + R(x)$. 

To solve the feasibility check problem, we turn the inner minimization problem into its Karush-Kuhn-Tucker (KKT) optimality condition:
\bsq
\label{eq:FC-KKT}
\begin{align}
   & 1-\pi -\theta =0, \label{eq:FC-KKT-1}\\
   & P^{\top}\pi = 0, \label{eq:FC-KKT-2}\\
   &  0 \le \pi \perp (Py+s-Q(x)\xi-R(x)) \ge 0 ,\label{eq:FC-KKT-3}\\
   &  0 \le \theta \perp s \ge 0 , \label{eq:FC-KKT-4}
\end{align}
\esq
% \bsq
% \label{eq:dual-FC}
% \begin{align}
%     \max_{\xi, \pi} ~ & (Q(x)\xi+R(x))^{\top} \pi \\
%     \mbox{s.t.}~& \eqref{eq:uncertainty-set}, P^{\top}\pi =0, 0 \le \pi \le 1
% \end{align}
% \esq
where $\pi$ and $\theta$ are the dual variables. Substituting \eqref{eq:FC-KKT} into the \textbf{FC} problem and linearizing the complementary slackness conditions \eqref{eq:FC-KKT-3}-\eqref{eq:FC-KKT-4} using the Big-M method, we get a mixed-integer linear program (MILP) that can be solved efficiently by commercial solvers. If the first-stage decision $x$ passes the feasibility check, i.e., the optimal solution of   \eqref{eq:FC} is $s^*=0$, then we move on to solve the subproblem:
\begin{subequations}
\label{eq:SP}
\begin{align}
\textbf{SP}: ~\max_\xi~ & \min_y E(x)^\top y, \\
\text{s.t.}~ & \eqref{eq:uncertainty-set}, P y  \geq Q(x) \xi + R(x).
\end{align}
\end{subequations}

Similarly, we turn the inner minimization problem into its KKT condition:
\bsq
\label{eq:SP-KKT}
\begin{align}
    & P^{\top}\pi = E(x), \\
    & 0 \le \pi \perp (Py-Q(x)\xi -R(x)) \ge 0.
\end{align}
\esq
% \bsq
% \label{eq:dual-SP}
% \begin{align}
%     \max_{\xi, \pi} ~ & (Q(x)\xi+R(x))^{\top} \pi \\
%     \mbox{s.t.}~& \eqref{eq:uncertainty-set}, P^{\top}\pi =E(x), \pi \ge 1
% \end{align}
% \esq

Again, substituting \eqref{eq:SP-KKT} into the \textbf{SP} problem \eqref{eq:SP}, we get an MILP. Solving the \textbf{FC} or \textbf{SP} problem, we can obtain a worst-case scenario $\xi^*$. 

\subsubsection{Representation of Worst-Case Scenario} Conventionally, we directly return the worst-case scenario $\xi^*$ to the master problem. However, since \eqref{eq:uncertainty-set} is a decision-dependent uncertainty set, it has been revealed that this conventional method may lead to over-conservativeness or failure of convergence \cite{tan2025adjustable}. To address this issue, we propose an innovative representation of the worst-case scenario: According to \cite{chen2022robust}, $\xi$ always resides at a vertex of the uncertainty set \eqref{eq:uncertainty-set}. Therefore, it can be represented by
\begin{align}\label{eq:9}
    \xi_{it}=\sum_{k \in \mathcal{K}} z_{tk}\left((1-v_{it})\xi_{itk}^{-}+v_{it}\xi_{itk}^{+}\right),
\end{align}
where $v_{it}$ is a binary variable. Moreover, to reduce conservativeness, we add the following constraints to the uncertainty set, where $\Gamma_T$ and $\Gamma_S$ are uncertainty budgets.
\begin{subequations}
\begin{align}
   \sum \nolimits_i |2v_{it}-1| \le \Gamma_S, \forall t, \\
    \sum \nolimits_t |2v_{it}-1| \le \Gamma_T, \forall i.
\end{align}
\end{subequations}
For each worst-case scenario $\xi^*$, we can determine the corresponding $v^*$ based on \eqref{eq:9}. Then, instead of passing $\xi^*$ directly to the master problem, we return the following constraints:
\begin{subequations}
\label{eq:WS}
\begin{align}
    & \xi_{itm}=\sum_{k \in \mathcal{K}} z_{tk}\left((1-v_{itm}^*)\xi_{itk}^{-}+v_{itm}^*\xi_{itk}^{+}\right),  \forall i \in \mathcal{I}, \forall t \in \mathcal{T}, \label{eq:WS-1}\\
     & P y_m \geq Q(x) \xi_m + R(x). \label{eq:WS-2}
\end{align}
\end{subequations}
where the index $m$ indicates the worst-case scenario in the $m$-th iteration. With \eqref{eq:WS}, when the first-stage decision $c_t^{TOU}$ changes, the $z_{tk}$ will change accordingly based on  \eqref{eq:uncertainty-set-2}--\eqref{eq:uncertainty-set-4}. Thus, the scenario $\xi_m = (\xi_{i t m})_{i \in \mathcal{I}, t \in \mathcal{T}}$ will be mapped to the vertex of the active respective range in \eqref{eq:uncertainty-set-origin} based on \eqref{eq:WS-1}. 

\subsubsection{Master Problem}
\label{subsubsec}
Given a set of vectors $\mathcal{V} = \{v_m^* = (v_{i t m}^*; i \in \mathcal{I}, t \in \mathcal{T}) \in [0, 1]^{IT}: m \in \mathcal{M}\}$ returned from previous iterations from \textbf{FC} or \textbf{SP}, the master problem is formulated by:
\begin{subequations}
\label{eq:master}
\begin{align}
\textbf{MP}:~\min_{x, z, \eta, \xi, y}~ & C^\top x + \eta, \label{eq:master-1}\\
\text{s.t.}~ & A x \geq B, \eqref{eq:uncertainty-set-2}\text{--}\eqref{eq:uncertainty-set-4}, \label{eq:master-2}\\
& \eta \geq E(x)^\top y_m, \forall m \in \mathcal{M}, \label{eq:master-3}\\
& \eqref{eq:WS}, \forall m \in \mathcal{M}. \label{eq:master-4}
\end{align}
\end{subequations}

However, the master problem cannot be solved directly due to the nonlinear constraint \eqref{eq:master-3} with a bilinear term $E(x)^{\top}y_m$; more specifically, the term $c_t^{TOU} l_{it}$. According to \eqref{eq:second-stage-1} and \eqref{eq:WS-1}, we have 
%l_{i t} = L_{i t}+ \xi_{i t}
%(c_t^{TOU} / c^{REF} - 1), \forall i \in \mathcal{I},
\begin{align}
    l_{i tm} = L_{i t} + \sum_{k \in \mathcal{K}} z_{tk}\phi_{km}(c_t^{TOU}/c^{REF} - 1),
\end{align}
where
\begin{align}
    \phi_{km}=L_{it}  \left((1-v_{itm}^*)\xi_{itk}^{-}+v_{itm}^*\xi_{itk}^{+}\right).
\end{align}
Therefore, the term $c_t^{TOU} l_{it}$ can be equivalently represented by an expression of the term $(c_t^{TOU})^2 z_{tk}$, where $z_{tk}$ is a binary variable. This can be turned into solvable constraints by replacing the term $(c_t^{TOU})^2 z_{tk}$ in the objective function by $\omega_{tk}$ and adding the following constraints with a large enough constant $M$:
\bsq\label{eq:transform}
\begin{align}
   & \omega_{tk} \ge 0, \\
   & (c_t^{TOU})^2 \le \omega_{tk} + M (1-z_{tk}).
\end{align}
\esq
When $z_{tk}=0$, we have $ \omega_{tk} \ge (c_t^{TOU})^2 $, so minimization ends up at $ \omega_{tk} = (c_t^{TOU})^2 $. When $z_{tk}=1$, we have $\omega_{tk} \ge 0$ and $(c_t^{TOU})^2 \le \omega_{tk} + M$. Since $M$ is a large enough constant, the latter constraint always holds. Therefore, minimization ends up at $ \omega_{tk} = 0$.

\subsubsection{Overall algorithm}
The overall procedure of the proposed improved C\&CG algorithm is presented in Algorithm 1. The convergence and optimality of Algorithm 1 are stated in Proposition~\ref{prop:algorithm}, whose proof is given in Appendix~\ref{apen-algorithm}.

\begin{algorithm}
\normalsize
\label{Ag:ICCG2}
\caption{{\bf : Improved C\&CG Algorithm}}
\begin{algorithmic}[1]
\STATE \emph{Initiation}:  Error tolerance $\tau>0$; $N=1$; Let $\mathcal{V}$ be an empty set and assign a large value to $UB_0$.
 
\STATE Solve the $\textbf{MP}$ problem \eqref{eq:master}, using the technique in  \eqref{eq:transform} to address the term $(c_t^{TOU})^2 z_{tk}$. Let $x^{N*}$ be the optimal solution and $LB_N$ be the optimal value.

\STATE Solve the \textbf{FC} problem \eqref{eq:FC} with $x^{N*}$. If the optimal solution $s^*=0$, go to step 4; otherwise, let $\xi^{N*}$ be the worst-case scenario and derive the corresponding $v^{N*}$ by \eqref{eq:9}, let $UB_N=UB_{N-1}$, and go to Step 5.

\STATE Solve the $\textbf{SP}$ problem \eqref{eq:SP} with $x^{N*}$. Let $(\xi^{N*}, y^{N*})$ be the optimal solution and $UB_N=C^\top x^{N*}+E(x^{N*})^{\top}y^{N*}$, and derive the corresponding $v^{N*}$ by \eqref{eq:9}. Go to Step 5.

\STATE if $|UB_N-LB_N|\le \tau$, terminate and output $x^{N*}$. Otherwise, add $v^{N*}$ into $\mathcal{V}$ and let $N=N+1$; go to Step 2.

\end{algorithmic}
\end{algorithm} 

\begin{proposition}
\label{prop:algorithm}
    Let $\overline{N} = 2^{I T} + 1$, where $I$ is the number of nodes and $T$ is the number of periods. If error tolerance $\tau = 0$, then Algorithm 1 converges within $\overline{N}$ iterations and outputs the optimal solution of problem \eqref{compact}.
\end{proposition}

\section{Numerical Experiments}
\label{sec:case}
In this section, we test the performance of the proposed model and solution method using the IEEE-33 bus system, where 3 additional generators are connected to nodes 2, 3, and 6, respectively. All the simulations are conducted on a laptop with Intel(R) Core(TM) Ultra 9 185H 2.50 GHz processor and 32 GB RAM, using MATLAB R2024A with GUROBI 11.03.

\subsection{Simulation Settings}
\begin{figure}[ht]
\centering
\includegraphics[width=0.85\columnwidth]{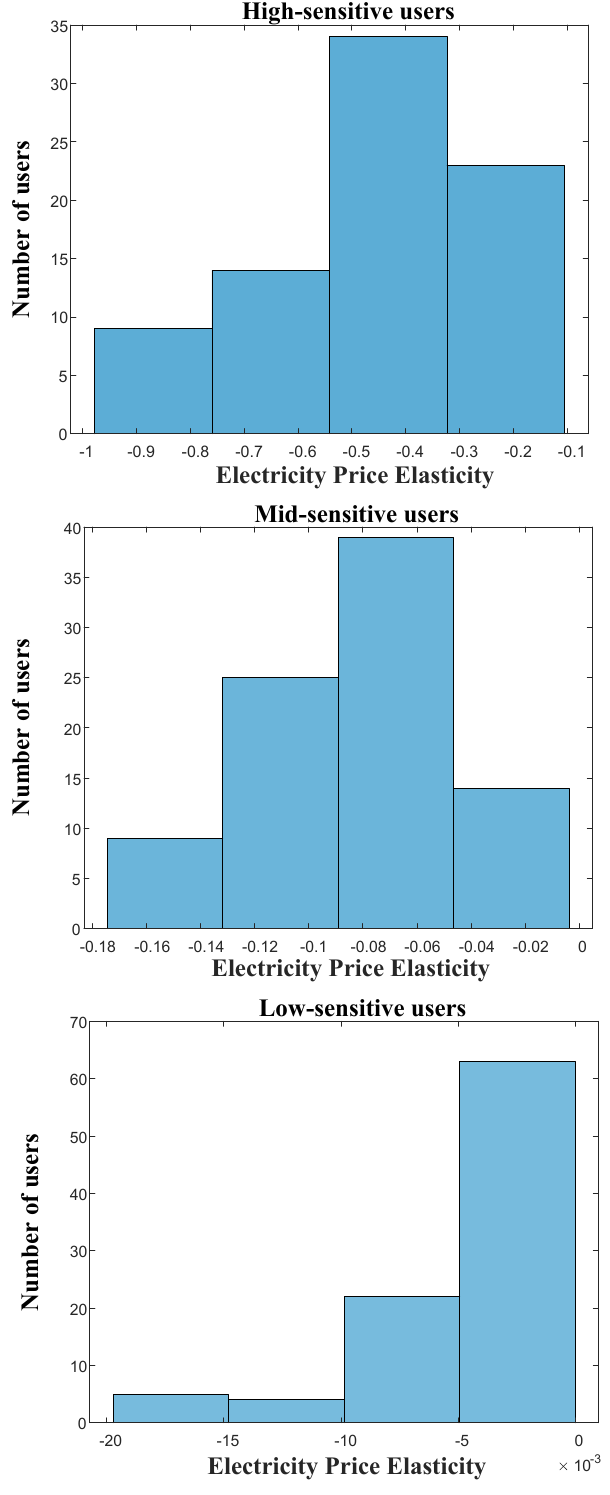}
\caption{Up: High-sensitive user distribution at electricity price ratio 0.3; Middle: Medium-sensitive user distribution at electricity price ratio 3; Down: Low-sensitive user distribution at electricity price ratio 16.}
\label{fig:London}
\end{figure}
First, we use the real-world energy consumption data from the Low Carbon London project \cite{LCL} to form the uncertainty set. The dataset contains millions pieces of data. However, the amount of useful data is much less than the amount of the entire full dataset, because most of the users were using fixed electricity and during the period of data, ToU was only applied in 2013. For these reasons, the selected data only accounts for a little of the entire data. Also, the referenced electricity price is calculated by average real-time electricity price in 2024 in AECO cite from PJM data miner \cite{pjm_data_miner}. We consider three types of users: high-sensitive users, mid-sensitive users, and low-sensitive users, categorized based on the range of their electricity price elasticity. According to the definition of the uncertainty set \eqref{eq:uncertainty-set-origin}, the elasticity is mainly influenced by the ratio between TOU price $c_t^{TOU}$ and the reference price $C_t^{REF}$. Therefore, here, we divide the range of TOU-to-reference ratio into five intervals, i.e., $[0.00, 0.25]$, $[0.25, 0.50]$, $[0.50, 1.00]$, $[1.00, 4.00]$, $[4.00, 16.00]$, respectively. For each type of users, we characterize their variation range of electricity price elasticity for each of these intervals. Fig. \ref{fig:London} shows three representative results. For example, Fig. \ref{fig:London}(a) shows the electricity price elasticity of high-sensitive users under $c_t^{TOU}/C_t^{REF}=0.3$. As we can see, the elasticity ranges from $-0.97$ to $-0.11$, which is chosen as the parameters in the uncertainty set. Similarly, we can determine the parameters for all types of users under all TOU-to-reference ratios, which gives the uncertainty set.

%{\color{red} Fig.\ref{fig:London} is a combination of Examples how price elasticity changing with TOU Ratio. As for the sensitivity of users toward TOU, 3 kinds of users including high sensitive users, mid sensitive users and low sensitive user, which are differentiated by their deviation. The range of TOU ratios are divided by 4 intervals originally. Therefore, there are 12 scenarios and 3 representative scenarios are selected as sub-figures. As it is indicated in the caption, the distributions on the x-axis show the range of price elasticity in the certain scenarios. In total, we can get figures including all 4 intervals, but there was not any data between [0.5,1]. Therefore, average of upper and lower boundary values of next bars are calculated as the the bounds of bar between [0.5,1]}

% \begin{figure}[ht]
% \centering
% \includegraphics[width=0.85\columnwidth]{figures/midrange.pdf}
% \caption{Mid-sensitive Users Price Elasticity Distribution Bars in 5 intervals }
% \label{fig:mid}
% \end{figure}
% \nocite{*}
\begin{figure}[ht]
\centering
\includegraphics[width=1.0\columnwidth]{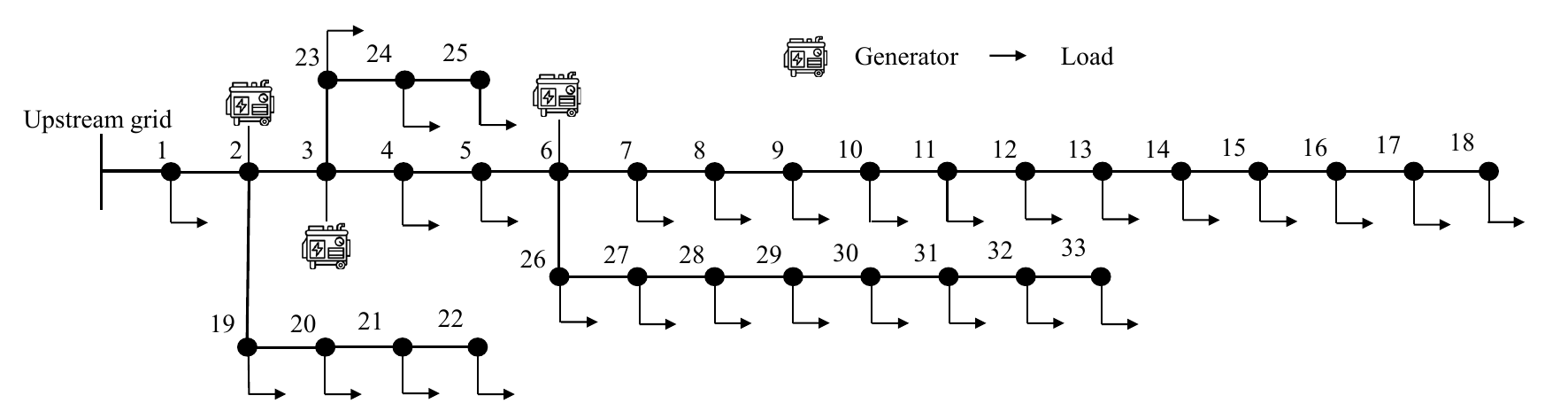}
\caption{The topology of the IEEE-33 bus system.}
\label{fig:system}
\end{figure}
\subsection{Benchmark Results}
\begin{figure}[ht]
\centering
\includegraphics[width=0.85\columnwidth]{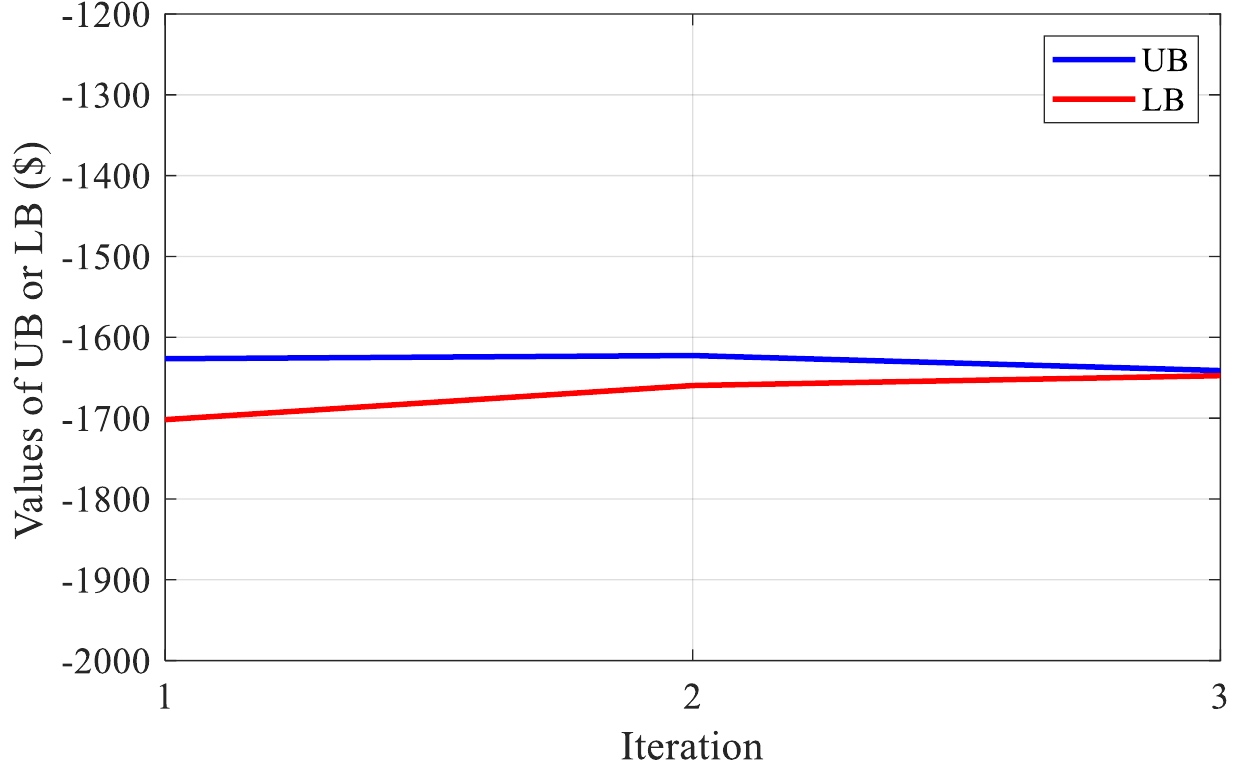}
\caption{Change of UB and LB during the iterations of Algorithm 1.}
\label{fig:linearcon}
\end{figure}

\begin{figure}[ht]
\centering
\includegraphics[width=0.85\columnwidth]{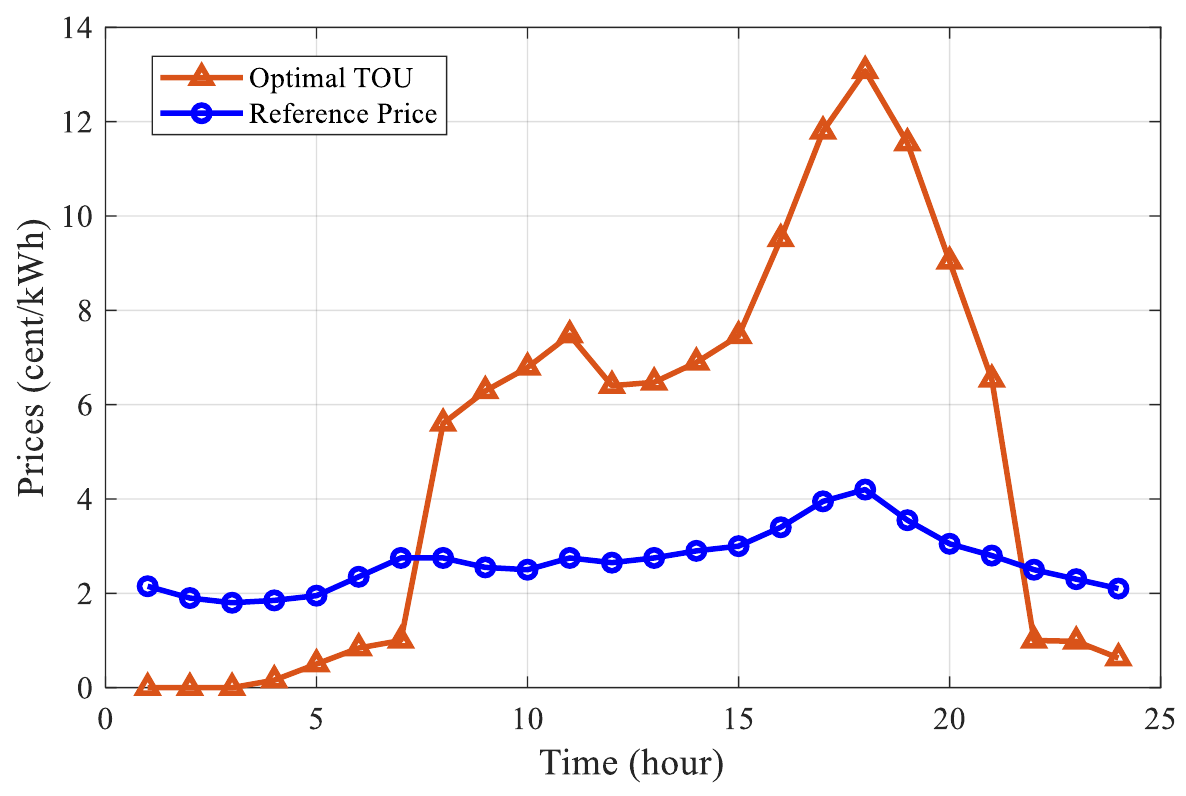}
\caption{Optimal TOU and reference prices of a day.}
\label{fig:prices}
\end{figure}

\begin{figure}[ht]
\centering
\includegraphics[width=0.85\columnwidth]{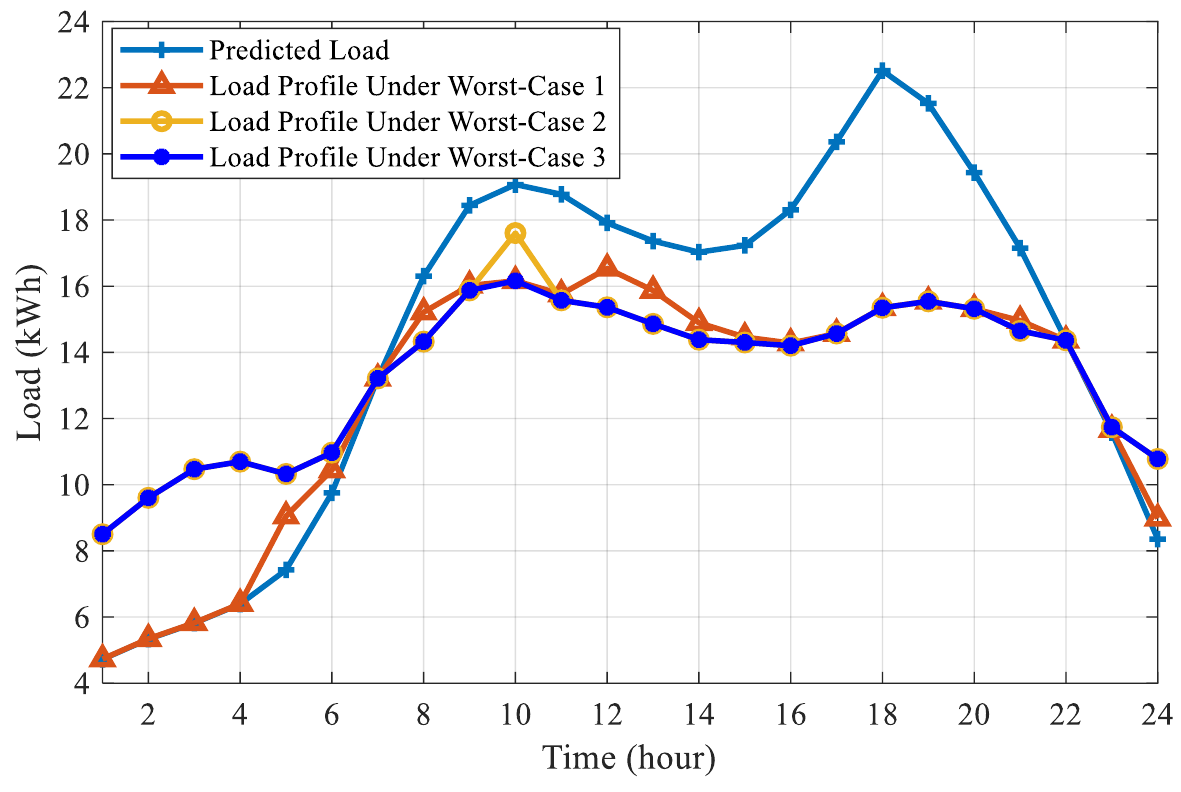}
\caption{Predicted and worst-case load profiles.}
\label{fig:load}
\end{figure}

Using the uncertainty set generated in the last subsection and the proposed model and method, we can get the results for the IEEE-33 bus systems. The topology of the system is shown in Fig. \ref{fig:system}. The proposed algorithm converges in 3 iterations, which takes 115.44 seconds. The time needed is acceptable for the operation of virtual power plants. The change of $LB$ and $UB$, defined in Algorithm 1, during the iteration processes is shown in Fig. \ref{fig:linearcon}; and they converge to -\$1645. Particularly, the negative value implies that the system is gaining profits by selling electricity. The optimal TOU prices obtained and the reference prices are shown in Fig. \ref{fig:prices}. The average predicted load $L_{it}$ of all buses and the average worst-case load profiles are given in Fig. \ref{fig:load}. As we can see, the optimal TOU price curve follows a similar patten to the load demand curve in Fig. \ref{fig:load}, i.e., the price is relatively high during periods with large demand while the price is relatively low when the load demand is small. This indicates that the TOU prices set by the proposed method can more effectively reflect the change of needs in the market than the reference price curve (blue curve in Fig. \ref{fig:prices}), which is almost flat. Moreover, the three load profiles under worst-cases in Fig. \ref{fig:load} are all smoother than the predicted load profile. This indicates that, by realizing and leveraging the demand flexibility through TOU prices, we can achieve peak shaving and valley filling, and thus, reduces the electricity bills.

%The optimization results are listed in the following figures.As is shown in fig. \ref {fig:linearcon}, the optimization process ends in 4 iterations in 60.1s, which is suitable for hourly ahead planning. The optimal result converges at -164.5. The negative result means that the system is gaining profits by selling electricity as the optimization goal is the cost.

\subsection{Necessity of the proposed methods}
\begin{figure}[ht]
\centering
\includegraphics[width=0.85\columnwidth]{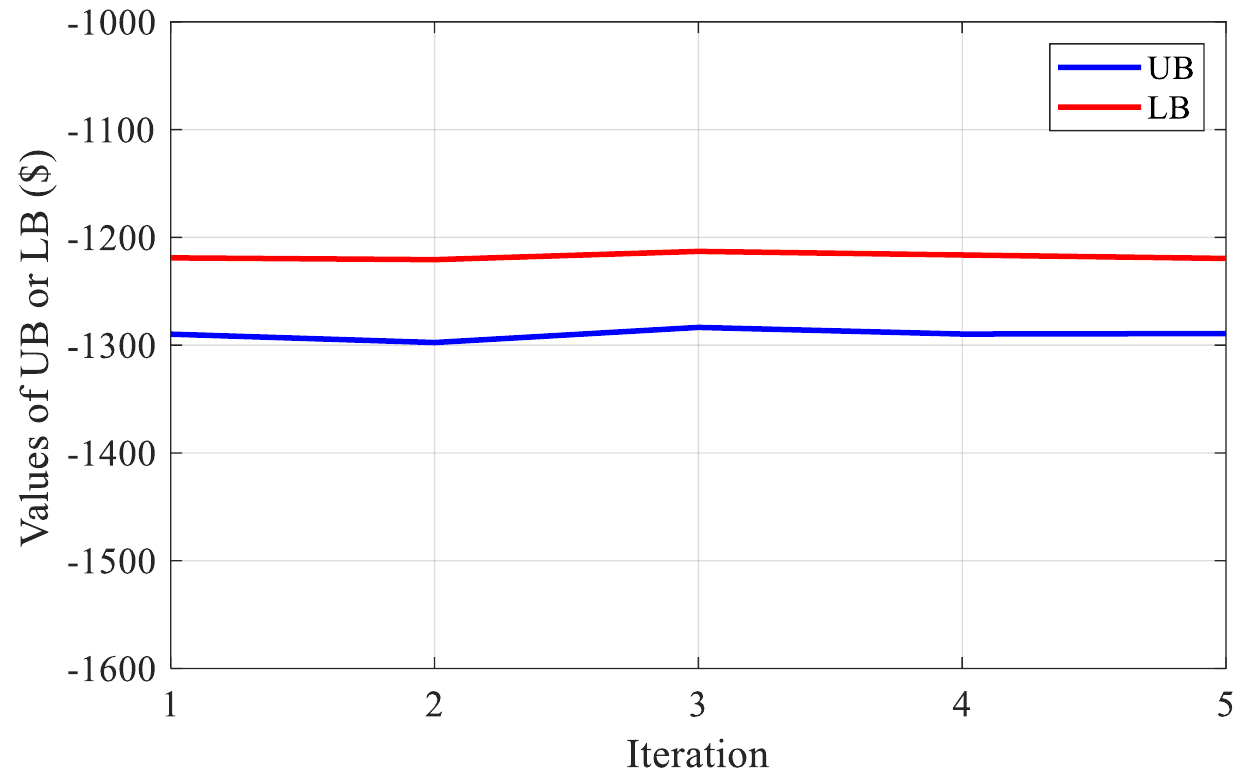}
\caption{Result of traditional C\&CG Algorithm.}
\label{CCGC}
\end{figure}

First, to show the necessity of the proposed RO algorithm, a traditional C\&CG is applied to optimize the model for comparison. More specifically, instead of returning the $v_{it}^*$ to the master problem as in the proposed algorithm, the traditional C\&CG algorithm returns the scenario $\xi^*$ directly to master problem. The result of the traditional C\&CG algorithm is shown in Fig. \ref{CCGC}. As we can see, the traditional C\&CG algorithm fails to converge, with the upper and lower bounds remain nearly unchanged. In addition, the value of LB is even larger than the value of UB, which is probably caused by the inaccurate estimation of $\xi$ by ignoring its decision-dependency in the scenarios. This shows the necessity of a new RO algorithm to address the DDU.

%As a comparison, a traditional C\&CG is applied to optimize the model. The fig \ref {CCGC} indicates that the lower bound from master problem also approaches -164.5, which is close to the result of proposed method. However, when it comes to the upper bound, the outcome of subproblem appears infeasible or unbounded,, which might result from the change in uncertainty set. Also, the optimzation process took more than 50 iterations and it was convinced not to converge. 

\begin{figure}[ht]
\centering
\includegraphics[width=0.85\columnwidth]{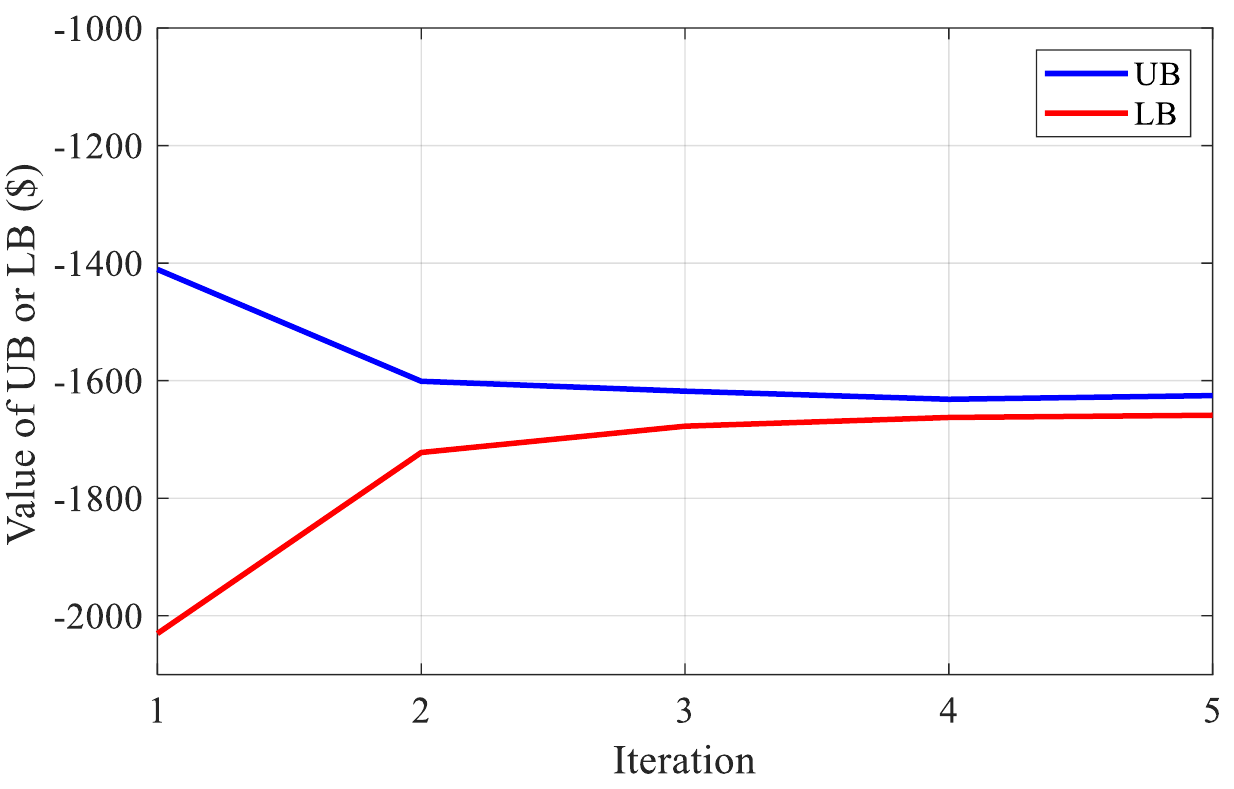}
\caption{Result of Nonlinear Model Convergence Figure}
\label{fig:NLC}
\end{figure}

Next, to show the accuracy of the proposed transformation method in Section \ref{subsubsec}, we compare the result using the proposed transformation method and the one using nonlinear models. The latter is shown in Fig. \ref{fig:NLC}. The nonlinear model takes 249.49 seconds to approximately converge, with the lower bound equaling -\$1662.6 and the upper bound equaling -\$1631.7, which are slightly larger than the -\$1645 obtained by the proposed algorithm. This reveals that the original nonlinear model could result in a suboptimal result.

\subsection{Sensitivity Analysis}
Further, we analyze the impact of different factors to better show the effectiveness of the proposed method.

\subsubsection{Fixed TOU price}
We test the influence of TOU prices. Instead of allowing the VPP operator to optimize the TOU prices, we use fixed TOU prices (constant $c_t^{TOU},\forall t$). We let $c_t^{TOU}$ be 1, 2, and 5 times of $C_t^{REF}$, respectively. The results are summarized in TABLE \ref{tab:comparison}. We can find that, with the optimal TOU prices derived by the proposed method, the VPP operator can generate the highest profit (the lowest objective value). This demonstrates the advantages of treating electricity prices as decision variables.

\begin{table}[H]
\renewcommand{\arraystretch}{1.2}
\caption{Results under different TOU prices}
\label{tab:comparison}
\centering
\footnotesize
\vspace{0.5em}
\begin{tabular}{lcccc}
\hline
$c_t^{TOU}$ & \textbf{Optimal} & $C_t^{REF}$ &  $2C_t^{REF}$  &  $5C_t^{REF}$   \\
\hline
Obj (\$) & -1645 &  73.65 & -260.8 & -1178.3 \\
\hline
\end{tabular}
\end{table}

\subsubsection{Uncertainty budget}
The level of uncertainty is also another important influencing factor, which is controlled by the uncertainty budgets $\Gamma_T$ and $\Gamma_S$. In the benchmark case, we let $\Gamma_T=24$ and $\Gamma_S=33$. Here, we test the impact of uncertainty level by changing these two parameters, with the results summarized in TABLE \ref{tab:comparison2}. As shown, the VPP operator's profit enhances with both reduced $\Gamma_T$ and $\Gamma_S$. However, smaller uncertainty budgets also result in more number of iterations and longer computation time. Still, the time needed is acceptable for the operation of VPP.

\begin{table}[H]
\renewcommand{\arraystretch}{1.2}
\caption{Results under different uncertainty levels}
\label{tab:comparison2}
\centering
\footnotesize
\vspace{0.5em}
\begin{tabular}{lcccc}
\hline
$(\Gamma_T, \Gamma_S)$ & $(24,33)$ & $(20,33)$ &  $(24, 25)$  &  $(20, 25)$   \\
\hline
Obj (\$) & -1645 &  -1648 & -1667 & -1668 \\
Time (s) & 115.44 & 287.95 & 276.36 & 283.02\\
No. of Iteration & 3 & 4 & 4 & 5\\
\hline
\end{tabular}
\end{table}

\section{Conclusion}
\label{sec:conclusion}
This paper considers the optimal operation of virtual power plants under DDU in the price elasticity of demand. We develop a two-stage robust VPP pricing model to determine TOU tariffs. The uncertain demand is influenced by the TOU pricing decisions. To solve this RO problem with DDU efficiently, we customize an improved C\&CG algorithm. This algorithm maps the previously selected worst-case scenarios to the vertices of the new uncertainty set, and can address a decision-dependent mixed-integer uncertainty set. The case studies reveal that compared to the traditional C\&CG algorithm, the proposed algorithm offers convergence guarantee under DDU. In the future, we may further explore methods to more precisely characterize the decision-dependency of uncertainty, as well as more effective RO algorithms for addressing DDU.

\bibliography{cite}

% Generated by IEEEtran.bst, version: 1.14 (2015/08/26)
\begin{thebibliography}{10}
\providecommand{\url}[1]{#1}
\csname url@samestyle\endcsname
\providecommand{\newblock}{\relax}
\providecommand{\bibinfo}[2]{#2}
\providecommand{\BIBentrySTDinterwordspacing}{\spaceskip=0pt\relax}
\providecommand{\BIBentryALTinterwordstretchfactor}{4}
\providecommand{\BIBentryALTinterwordspacing}{\spaceskip=\fontdimen2\font plus
\BIBentryALTinterwordstretchfactor\fontdimen3\font minus \fontdimen4\font\relax}
\providecommand{\BIBforeignlanguage}[2]{{%
\expandafter\ifx\csname l@#1\endcsname\relax
\typeout{** WARNING: IEEEtran.bst: No hyphenation pattern has been}%
\typeout{** loaded for the language `#1'. Using the pattern for}%
\typeout{** the default language instead.}%
\else
\language=\csname l@#1\endcsname
\fi
#2}}
\providecommand{\BIBdecl}{\relax}
\BIBdecl

\bibitem{IEA2025}
\BIBentryALTinterwordspacing
IEA, ``Global energy review 2025,'' IEA, Paris, France, Tech. Rep., 2025. [Online]. Available: \url{https://www.iea.org/reports/global-energy-review-2025}
\BIBentrySTDinterwordspacing

\bibitem{ruan2024VPP}
G.~Ruan, D.~Qiu, S.~Sivaranjani, A.~S. Awad, and G.~Strbac, ``Data-driven energy management of virtual power plants: A review,'' \emph{Advances in Applied Energy}, vol.~14, p. 100170, 2024.

\bibitem{ullah2019comprehensive}
Z.~Ullah, G.~Mokryani, F.~Campean, and Y.~F. Hu, ``Comprehensive review of {VPPs} planning, operation and scheduling considering the uncertainties related to renewable energy sources,'' \emph{IET Energy Systems Integration}, vol.~1, no.~3, pp. 147--157, 2019.

\bibitem{shinde2022multistage}
P.~Shinde, I.~Kouveliotis-Lysikatos, and M.~Amelin, ``Multistage stochastic programming for {VPP} trading in continuous intraday electricity markets,'' \emph{IEEE Transactions on Sustainable Energy}, vol.~13, no.~2, pp. 1037--1048, 2022.

\bibitem{Wulro}
X.~Wu, H.~Xiong, S.~Li, S.~Gan, C.~Hou, and Z.~Ding, ``Improved light robust optimization strategy for virtual power plant operations with fluctuating demand,'' \emph{IEEE Access}, vol.~11, pp. 53\,195--53\,206, 2023.

\bibitem{yu2023flexible}
S.~Yu, F.~Fang, and J.~Liu, ``Flexible operation of a {CHP-VPP} considering the coordination of supply and demand based on a strengthened distributionally robust optimization,'' \emph{IET Control Theory \& Applications}, vol.~17, no.~16, pp. 2146--2161, 2023.

\bibitem{cao2023distributionally}
J.~Cao, B.~Yang, C.~Y. Chung, Y.~Gong, and X.~Guan, ``Distributionally robust management of hybrid energy station under exogenous-endogenous uncertainties and bounded rationality,'' \emph{IEEE Transactions on Sustainable Energy}, vol.~15, no.~2, pp. 884--902, 2023.

\bibitem{xiang2023optimizing}
D.~Xiang, K.~He, G.~Zhang, S.~Zhang, Y.~Chen, L.~Zhao, H.~Liu, X.~Su, and H.~Zheng, ``Optimizing {VPP} operations: A novel robust demand-side management approach with energy storage system,'' in \emph{2023 IEEE International Conference on Energy Internet (ICEI)}.\hskip 1em plus 0.5em minus 0.4em\relax IEEE, 2023, pp. 45--50.

\bibitem{nemati2025assessingvaluerenewablebasedvpp}
\BIBentryALTinterwordspacing
H.~Nemati, I.~Egido, P.~S{\'a}nchez-Mart{\'\i}n, and {\'A}.~Ortega, ``Assessing value of renewable-based {VPP} versus electrical storage: Multi-market participation under different scheduling regimes and uncertainties,'' 2025. [Online]. Available: \url{https://arxiv.org/abs/2507.22496}
\BIBentrySTDinterwordspacing

\bibitem{bertsimas2012adaptive}
D.~Bertsimas, E.~Litvinov, X.~A. Sun, J.~Zhao, and T.~Zheng, ``Adaptive robust optimization for the security constrained unit commitment problem,'' \emph{IEEE Transactions Power Systems}, vol.~28, no.~1, pp. 52--63, Feb. 2013.

\bibitem{zeng2013solving}
B.~Zeng and L.~Zhao, ``Solving two-stage robust optimization problems using a column-and-constraint generation method,'' \emph{Operations Research Letters}, vol.~41, no.~5, pp. 457--461, 2013.

\bibitem{tan2025adjustable}
T.~Tan, M.~Yang, R.~Xie, Y.~Cao, and Y.~Chen, ``Adjustable robust optimization with decision-dependent uncertainty for power system problems: A review,'' \emph{Energy Conversion and Economics}, 2025.

\bibitem{zhang2021robust}
Y.~Zhang, F.~Liu, Z.~Wang, Y.~Su, W.~Wang, and S.~Feng, ``Robust scheduling of virtual power plant under exogenous and endogenous uncertainties,'' \emph{IEEE Transactions on Power Systems}, vol.~37, no.~2, pp. 1311--1325, 2021.

\bibitem{qi2023portfolio}
N.~Qi, L.~Cheng, H.~Li, Y.~Zhao, and H.~Tian, ``Portfolio optimization of generic energy storage-based virtual power plant under decision-dependent uncertainties,'' \emph{Journal of energy storage}, vol.~63, p. 107000, 2023.

\bibitem{hellemo2018decision}
L.~Hellemo, P.~I. Barton, and A.~Tomasgard, ``Decision-dependent probabilities in stochastic programs with recourse,'' \emph{Computational Management Science}, vol.~15, no.~3, pp. 369--395, 2018.

\bibitem{zhou2023urban}
K.~Zhou, N.~Peng, H.~Yin, and R.~Hu, ``Urban virtual power plant operation optimization with incentive-based demand response,'' \emph{Energy}, vol. 282, p. 128700, 2023.

\bibitem{mei2023optimal}
S.~Mei, Q.~Tan, Y.~Liu, A.~Trivedi, and D.~Srinivasan, ``Optimal bidding strategy for virtual power plant participating in combined electricity and ancillary services market considering dynamic demand response price and integrated consumption satisfaction,'' \emph{Energy}, vol. 284, p. 128592, 2023.

\bibitem{kong2023real}
X.~Kong, W.~Lu, J.~Wu, C.~Wang, X.~Zhao, W.~Hu, and Y.~Shen, ``Real-time pricing method for {VPP} demand response based on {PER-DDPG} algorithm,'' \emph{Energy}, vol. 271, p. 127036, 2023.

\bibitem{lu2025assessing}
G.~Lu, B.~Yuan, S.~Zhou, L.~Wei, and Z.~Wu, ``Assessing the effectiveness of time-of-use pricing design: Provincial evidence from {China},'' \emph{Energy Strategy Reviews}, vol.~60, p. 101780, 2025.

\bibitem{kirschen2004fundamentals}
D.~S. Kirschen and G.~Strbac, \emph{Fundamentals of power system economics}.\hskip 1em plus 0.5em minus 0.4em\relax Wiley, 2004.

\bibitem{chen2022robust}
Y.~Chen and W.~Wei, ``Robust generation dispatch with strategic renewable power curtailment and decision-dependent uncertainty,'' \emph{IEEE Transactions on Power Systems}, vol.~38, no.~5, pp. 4640--4654, 2022.

\bibitem{Xieload}
R.~Xie, P.~Pinson, Y.~Xu, and Y.~Chen, ``Robust generation dispatch with purchase of renewable power and load predictions,'' \emph{IEEE Transactions on Sustainable Energy}, vol.~15, no.~3, pp. 1486--1501, 2024.

\bibitem{tan2024robust}
T.~Tan, R.~Xie, X.~Xu, and Y.~Chen, ``A robust optimization method for power systems with decision-dependent uncertainty,'' \emph{Energy Conversion and Economics}, vol.~5, no.~3, pp. 133--145, 2024.

\bibitem{yang2025robust}
M.~Yang, R.~Xie, Y.~Zhang, and Y.~Chen, ``Robust microgrid dispatch with real-time energy sharing and endogenous uncertainty,'' \emph{IEEE Transactions on Smart Grid}, 2025.

\bibitem{zhang2022two}
Y.~Zhang, F.~Liu, Y.~Su, Y.~Chen, Z.~Wang, and J.~P. Catal{\~a}o, ``Two-stage robust optimization under decision dependent uncertainty,'' \emph{IEEE/CAA Journal of Automatica Sinica}, vol.~9, no.~7, pp. 1295--1306, 2022.

\bibitem{avraamidou2020adjustable}
S.~Avraamidou and E.~N. Pistikopoulos, ``Adjustable robust optimization through multi-parametric programming,'' \emph{Optimization Letters}, vol.~14, no.~4, pp. 873--887, 2020.

\bibitem{vayanos2025robust}
P.~Vayanos, A.~Georghiou, and H.~Yu, ``Robust optimization with decision-dependent information discovery,'' \emph{Management Science}, 2025.

\bibitem{zhang2025generic}
Y.~Zhang, Y.~Su, and F.~Liu, ``On decision-dependent uncertainties in power systems with high-share renewables,'' \emph{Engineering}, 2025.

\bibitem{LCL}
\BIBentryALTinterwordspacing
U.~P. Networks, ``Smartmeter energy consumption data in london households,'' 2014, accessed: 2025-9-20. [Online]. Available: \url{https://innovation.ukpowernetworks.co.uk/projects/low-carbon-london}
\BIBentrySTDinterwordspacing

\bibitem{pjm_data_miner}
\BIBentryALTinterwordspacing
{PJM}, ``{PJM} data miner 2,'' 2024, settlement verified hourly LMPs of PJM market. [Online]. Available: \url{https://dataminer2.pjm.com/list}
\BIBentrySTDinterwordspacing

\end{thebibliography}
\bibliographystyle{IEEEtran}

\appendices

\makeatletter
\@addtoreset{equation}{section}
\@addtoreset{theorem}{section}

\makeatother
\renewcommand{\theequation}{A.\arabic{equation}}
\renewcommand{\thetheorem}{A.\arabic{theorem}}
\section{Proof of Proposition \ref{prop:algorithm}}
\label{apen-algorithm}

Suppose $O^*$ is the optimal value of problem \eqref{compact} and the solution $(x^*, z^*, \xi^*)$ is optimal. We prove Proposition~\ref{prop:algorithm} by showing three claims:

1) Claim 1: $LB_N \leq O^* \leq UB_N$ for any $N$.

Proof of Claim 1: According to the master problem \eqref{eq:master}, 
\begin{subequations}
\label{eq:proof-LB}
\begin{align}
    \label{eq:proof-LB-1}
    LB_N = \min_{x, z}~ & C^\top x + \max_{\xi, v} \min_y E(x)^\top y ,\\
    \label{eq:proof-LB-2}
    \text{s.t.}~ & A x \geq B, \eqref{eq:uncertainty-set-2}\text{--}\eqref{eq:uncertainty-set-4}, P y \geq Q(x) \xi + R(x), \\
    \label{eq:proof-LB-3}
    & \xi_{it}=\sum_{k \in \mathcal{K}} z_{tk}\left((1-v_{it})\xi_{itk}^{-}+v_{it}\xi_{itk}^{+}\right), \\
    \label{eq:proof-LB-4}
    & v \in \{v^{1*}, v^{2*}, \dots, v^{N-1*}\}.
\end{align}
\end{subequations}

Because
\begin{subequations}
\begin{align}
    O^* = \min_{x, z}~ & C^\top x + \max_{\xi, v} \min_y E(x)^\top y ,\\
    \text{s.t.}~ & \eqref{eq:proof-LB-2}, \eqref{eq:proof-LB-3}, v \in \mathcal{V},
\end{align}
and $\{v^{1*}, v^{2*}, \dots, v^{N-1*}\} \subset \mathcal{V}$, problem \eqref{eq:master} is a relaxation of 
\end{subequations} problem \eqref{compact}, which implies $LB_N \leq O^*$.

According to the subproblem \eqref{eq:SP},
\begin{subequations}
\label{eq:proof-UB}
\begin{align}
    UB_N = \min_{x, z}~ & C^\top x + \max_\xi \min_y E(x)^\top y ,\\
\text{s.t.}~ & A x \geq B, \eqref{eq:uncertainty-set}, P y \geq Q(x) \xi + R(x), \\
& x = x^{N*},
\end{align}    
\end{subequations}
so $O^* \leq UB_N$.

2) Claim 2: If $N_1 < N_2$ and Algorithm 1 does not converge after $N_2$ iterations, then $v^{N_1*} \neq v^{N_2*}$.

Proof of Claim 2: We prove it by contradiction. Suppose $v^{N_1*} = v^{N_2*}$, then $v^{N_2*} \in \{v^{1*}, v^{2*}, \dots, v^{N_2-1*}\}$. Using the optimal solution $x^{N_2*}$ of the master problem \eqref{eq:master} in the $N_2$-th iteration and the equation in \eqref{eq:proof-LB}, we have
\begin{subequations}
\begin{align}
    LB_{N_2} = \min_{x, z}~ & C^\top x + \max_{\xi, v} \min_y E(x)^\top y ,\\
    \text{s.t.}~ & \eqref{eq:proof-LB-2}, \eqref{eq:proof-LB-3}, \\
    & v \in \{v^{1*}, v^{2*}, \dots, v^{N_2-1*}\}, x = x^{N_2*}.
\end{align}
\end{subequations}
Substituting $\{v^{1*}, v^{2*}, \dots, v^{N_2-1*}\}$ by $\{ v^{N_2*} \}$, we have
\begin{subequations}
\begin{align}
    LB_{N_2} \geq \min_{x, z}~ & C^\top x + \max_{\xi, v} \min_y E(x)^\top y ,\\
    \text{s.t.}~ & \eqref{eq:proof-LB-2}, \eqref{eq:proof-LB-3}, \\
    & v = v^{N_2*}, x = x^{N_2*},
\end{align}
\end{subequations}
where the right-hand side equals $UB_{N_2}$ according to \eqref{eq:proof-UB} and the optimality of $v^{N_2*}$ in the subproblem \eqref{eq:SP}. Thus, $LB_{N_2} \geq UB_{N_2}$. Combining it with Claim 1, we have $UB_{N_2} = LB_{N_2}$ and Algorithm 1 converges after $N_2$ iterations, which is a contradiction.

3) Claim 3: Algorithm 1 converges within $\overline{N}$ iterations.

Proof of Claim 3: For any iteration $N$, the worst-case scenario $\xi^{N*}$ always resides at a vertex of the uncertainty set \eqref{eq:uncertainty-set} \cite{chen2022robust}. Therefore, $v^{N*}$ can be achieved at a vertex of the set $\mathcal{V} = [0, 1]^{I T}$. By Claim 2, a vertex of $\mathcal{V}$ cannot be found twice unless Algorithm 1 converges. Therefore, the number of iterations cannot exceed the number of vertices of $\mathcal{V}$ plus 1, which is $\overline{N} = 2^{I T} + 1$. 

According to Claim 3, we can assume Algorithm 1 converges after the $N_0$-th iteration, where $N_0 \leq \overline{N}$. By the convergence criterion, we have $UB_{N_0} = LB_{N_0}$. Thus, Claim 1 implies $LB_{N_0} = O^* = UB_{N_0}$, which proves Algorithm 1 finds the optimal solution. This completes the proof.
% \section{TOU Ratio Result}
% \begin{figure}[ht]
% \centering
% \includegraphics[width=1.0\columnwidth]{figures/CCGTOU.pdf}
% \caption{The TOU Ratio of C\&CG Algorithm}
% \label{fig:CCGTOU}
% \end{figure}
% \begin{figure}[ht]
% \centering
% \includegraphics[width=1.0\columnwidth]{figures/CTOURATIO.pdf}
% \caption{The TOU Ratio of Constant TOU Ratio}
% \label{fig:CTOU}
% \end{figure}
% \begin{figure}[ht]
% \centering
% \includegraphics[width=1.0\columnwidth]{figures/CDTOU.pdf}
% \caption{The TOU Ratio of Constant Elasticity Distribution}
% \label{fig:CDTOU}
% \end{figure}
% \begin{figure}[ht]
% \centering
% \includegraphics[width=1.0\columnwidth]{figures/CE0.1TOU.pdf}
% \caption{The TOU Ratio of Constant Elasticity value}
% \label{fig:CETOU}
% \end{figure}
\end{document}